# A Risk-Based Probabilistic Transient Stability Approach for Ranking of Circuit Breakers in a Power System

**Umair Shahzad**

School of Computer Science and Engineering, University of Sunderland,

United Kingdom of Great Britain and Northern Ireland

Email: umair.shahzad@sunderland.ac.uk

**Abstract**

Power systems are getting more complex than ever and are consequently operating close to their limit of stability. Moreover, with the increasing demand of renewable wind generation, and the requirement to maintain a secure power system, the importance of transient stability cannot be overestimated. Current deterministic industry practices of transient stability assessment ignore the probability of variables involved. With increasing system uncertainties and widespread electricity market deregulation, there is a strong inevitability to incorporate probabilistic transient stability analysis. Circuit breakers play a critical role in fault clearing and consequently in determining the system transient stability. It is important that they undergo timely and appropriate maintenance procedures based on some criterion. Considering the need of incorporating risk in modern power systems, this paper proposes a risk-based probabilistic transient stability approach for ranking of circuit breakers in a power system. A novel priority index was proposed to rank the circuit breakers based on the system transient stability risk. DIgSILENT PowerFactory software was used to conduct the required simulations on IEEE 14 bus system. The proposed risk-based framework was deemed to be efficient in identification of the circuit breakers based on their priority rank index which can aid in power system planning process.

## List of Abbreviations

| Abbreviation | Meaning |
|---|---|
| CB | Circuit Breaker |
| CCT | Critical Clearing Time |
| DER | Distributed Energy Resources |
| DFIG | Doubly Fed Induction Generator |
| DSA | Dynamic Security Assessment |
| FCT | Fault Clearing Time |
| IoT | Internet of Things |
| LG | Single Line to Ground |
| LL | Line to Line |
| LLG | Line to Line to Ground |
| LLL | Line to Line to Line |
| MC | Monte Carlo |
| PDF | Probabilistic Density Function |
| PMF | Probabilistic Mass Function |
| PTS | Probabilistic Transient Stability |
| PV | Photovoltaic |
| RMS | Root Mean Squared |
| SMES | Superconducting Magnetic Energy Storage |
| SMIB | Single Machine Infinite Bus |

## 1. Introduction

Driven by various techno-economic and environmental factors, the electric energy industry is anticipated to undergo a paradigm shift, with a significantly augmented level of renewables, especially, wind and solar power sources, gradually replacing conventional power production sources (coal, diesel, natural gas, etc.). This increasing demand of large-scale wind integration in the conventional power system, along with the inherent and external uncertainties of the system, brings a lot of challenges [1-2]. One of them is the power system transient stability. Power systems are regularly exposed to unanticipated faults. Such faults can cause transient instability and can consequently lead to prevalent outages [3]. To preserve system security, system operators and planners perform analysis to make critical operating and planning decisions that will ensure safe operation of the power system after the occurrence such faults. The current general practice, within the power industry, is to use the deterministic approaches, with significant safety margins, to cover all possible uncertainties. With the adoption of a deterministic criterion for system security, power systems generally operate with a large security margin. Usually, these deterministic criteria provide safe, but conservative limits for system operating conditions. The most crucial security criterion is the ($N-1$) security criterion that guarantees safe operation of the power system, after the failure of a single element of the system, where $N$ is the total number of system components [3-5].

In the past several years, there has been a significant increase in connections of intermittent and stochastic, power electronics interfaced renewable energy generation sources. These uncertainties, coupled with load uncertainties, are becoming one of the vital characteristics of modern power systems. The transient stability assessment of such systems, using traditional deterministic methodology, is swiftly becoming inappropriate and thus, innovative probabilistic assessment approaches are desirable [4-5].

The current industry practices use the deterministic approach for transient stability assessment [6-7]. Although, the deterministic approaches result in highly secured power systems, but they do not consider the probability of operating conditions. Apart from the high cost due to conservative designs, the key drawback with the deterministic assessment techniques is that they consider all security problems to have equal risk [8]. Various literature (including research papers, technical reports, dissertations, and white papers) [7-18] mention that probabilistic risk-based transient stability, and incorporating risk in power planning procedures, is a future research area, and consequently, work needs to be done in this domain. Moreover, planning guides/manuals of various utilities [19-24] recommend using risk-based probabilistic approaches in the near future. Probabilistic assessment methods [25-26] can be helpful when uncertain parameters are included into system stability assessment. It is, thus, of great implication to propose a risk-based approach, for overcoming the inadequacies of the deterministic approach [5].

Circuit breakers (CBs) play an important role in power system transient stability and reducing the probability of cascading outages which lead to blackouts or major power disruption. A CB is an electrical switch designed to protect an electrical circuit from damage caused by overcurrent/overload or short circuit. Its basic function is to interrupt current flow after protective relays detect a fault. A CB is a critical component in assessing system risk. Its timely maintenance is important to achieve power system security. Fault Clearing Time (FCT) is a critical factor in assessing transient stability. The smaller this time is, the greater is the probability that the system will be transiently stable after occurrence of a fault. The Operating and tripping time of CBs determine the FCT. Critical Clearing Time (CCT) is the maximum time a fault can be applied without the power system losing stability. If the fault is cleared before the CCT, the system remains stable. If the fault is cleared after the CCT, generators lose synchronism and the system becomes unstable. The kinetic energy picked by the synchronous generator (SG) is directly proportional to fault duration. Therefore, faster FCT results in a greater stability margin. In short, CBs contribute to transient stability by isolating faults, responding rapidly, coordinating with other protective devices, and managing transient conditions effectively. This helps maintain the integrity of the electrical system during disturbances. Hence, CBs play a critical role in fault clearing and consequently in determining the system transient stability. Although, some research involving transient stability has been conducted on CB maintenance and replacement [27-37], and risk [38-42]; however, to the best of author's knowledge, no work ranks the CBs based on transient stability risk. Hence, this paper proposes a risk-based probabilistic transient stability approach for ranking of CBs in a power system.

The rest of the paper is organised as follows. Section 2 gives an overview of risk-based probabilistic transient stability. Sections 3 and 4 describe the mathematical formulation used for the proposed approach and the associated computation procedure, respectively. Sections 5 and 6 describe the case study, and results and discussion, respectively. Finally, Section 7 concludes the paper with suggestions for future work.

## 2. Risk-based Probabilistic Transient Stability: Background and Overview

This section will review some literature, pertinent to probabilistic transient stability (PTS). A significant amount of literature is available in the domain of PTS. [43] proposed a probabilistic framework, for power system transient stability assessment, with high renewable generation penetration. The presented framework enables comprehensive calculation of transient stability of power systems, with abridged inertia. A major drawback of the work is considering only three-phase line faults. [44] presented a study of the effects of some important power system parameters on transient stability. The parameters considered for this assessment include fault location, load increment, machine damping factor, FCT, and generator synchronous speed. The work considers only three-phase line faults. In [45], the abridged version of an altered Single Machine Infinite Bus (SMIB) system, with a doubly fed induction generator (DFIG)-based wind farm integration, is analysed, considering the transient features of the DFIG-based wind farm, in the diverse periods of a fault. The assessment specifies that the performance of SG can either be enhanced or depreciated with DFIG integration. Only a three-phase fault is considered on a specific line, which clears after a pre-selected time. [46] presented the results of a PTS assessment, conducted on the large-scale system of B.C. Hydro, including a generation rejection study, on the Peace system and a transfer limit study on the Columbia system. In this paper, B.C. Hydro's historical statistics on the probabilistic states of load level factor, fault type, fault location, fault clearing, and automatic reclosing were used in a Monte Carlo (MC) formulation to produce sample states for the case studies [5].

[47] illustrated the incorporation of probabilistic analysis in transient stability of a practical power system, by applying it to a multimachine configuration. [48] provided an analytical algorithm, based on transient energy margin, for the online PTS assessment of existing or forecasted operating conditions. [49] presented two approaches, for computing the probability of transient instability. These methods are based on Bayesian theory and Cartesian products. [50] presented a stochastic-based method, to assess the PTS index of the power system, including the wind farm and the Superconducting Magnetic Energy Storage (SMES). Uncertain factors include both sequence of disturbance in power grid and stochastic generation of the wind farm. [51] proposed a stochastic-based approach, to determine the PTS indices of a power system, incorporating wind farms. In this scenario, research was conducted on a hypothetical test system, considering the uncertainties of the factors, associated with power system operation, namely fault type, fault location, fault impedance, fault clearing process, system parameters, operating conditions, and high-speed reclosing. [52] illustrated the method of bisection to analytically evaluate the PTS indices. [53] provided a probabilistic approach to assess the transient stability of a wind farm. It extended and illustrated a rudimentary procedure, for calculating the probability of transient stability, for each transmission line, and for the overall system [5].

A PTS assessment method, based on MC simulation, was proposed in [54]. Two instability probability indices were also defined as indicators of the overall system stability and the severity of individual component fault. In [55], a probabilistic assessment technique, based on quasi-MC method, to analyse the transient stability of power system, incorporating wind power, was presented. Two indices were introduced, to assess the transient stability of power system. [56] proposed an analytical assessment method, for transient stability assessment of multimachine power systems, under stochastic continuous disturbances. In the suggested technique, a probability measure of transient stability was presented and analytically solved by stochastic averaging. [57] proposed a method, for obtaining PTS assessment, by using distribution functions, based on location, fault type, and sequence. [58] proposed the use of MC simulations, in the computation of probabilistic measures, for the transient stability problem. [59] suggested approximate methods, for evaluating probabilistic transient instability, and classifying critical stability areas, for system planning. In [60-61], an approach was developed, for obtaining a stability index, for individual lines, and for the overall system, for numerous fault types. The impact of clearing times and reclosing times was also investigated for critical lines. In [62], an approach was presented to assess the distribution of the probability of instability. In [63], conditional probabilities were used in the evaluation of probabilistic transient instability. [64] proposed a real-time approach for computing probabilistic CCT which is applicable to PTS assessment. The goal of the proposed approach was to offer a low computational burden and high accuracy to calculate the probabilistic density function (PDF) of CCT in two stages [5].

[65] proposed a method for power system PTS assessment, considering the wind farm uncertainties and correlations. Specifically, the inverse Nataf transformation based three-point estimation method, and the Cornish-Fisher expansion were combined, to deal with the uncertainties, and the

correlations amongst numerous wind farms. In [66], two-point estimate method was used, to determine the maximum relative rotor angles' probability distribution functions, for a given fault, with uncertain load demands and clearing time. A probabilistic approach, to assess the transient stability of power systems, with increased penetration of wind and photovoltaic (PV) generation, was presented in [67]. The impact on transient stability, due to the intermittent behaviour of Distributed Energy Resources (DERs), and their dynamic response, when a disturbance happens, was examined. An analytical approach, for probabilistic dynamic security assessment (DSA) of power systems, incorporating wind farms, was presented in [68]. The probability of transient stability, given a specific fault and uncertainties of output power of wind farm and load was analytically computed [5].

Recent advancements in CB risk assessment focus on enhancing reliability, performance, and safety through improved technologies and methodologies. Some key developments are as follows. Advanced predictive analytics using machine learning to forecast failures based on operational data is gaining traction. Numerous research highlights models that analyse historical breaker performance data to predict potential failures [69]. The integration of Internet of Things (IoT) technology in CBs enables real-time monitoring of parameters like temperature, current, and voltage. This real-time data enhances risk assessments and operational decisions [70]. Non-invasive diagnostic techniques, such as partial discharge and thermal imaging, are increasingly utilized for assessing the condition of CBs without interrupting service [71].

CB risk assessment, while helpful, does have several drawbacks and challenges. Some of the main issues are as follows. The increasing amount of data generated by smart monitoring systems can overwhelm analysts, making it difficult to extract actionable insights [72]. Predictive maintenance models can sometimes generate false positives leading to redundant maintenance actions that increase costs and downtime [73]. As technology evolves, keeping up with regulatory standards can be challenging. Inconsistent regulations across regions may complicate risk assessment practices [74].

## 3. Mathematical Formulation

Risk-based approach describes possibility of contingency by probability, and the corresponding impact (or consequence) by severity function. The product of this probability and associated severity is termed as risk [75-77]. CCT is the maximum clearing time before which the fault must be cleared to keep the system transiently stable [78]. Based on the conceptual framework of risk mentioned in [75-80], let $R_i$ be the transient instability risk for $i^{th}$ MC sample. Similarly, let $R_A$ be the average risk index for transient instability. Mathematically,

$$R_i = \Pr(U_i \cap F_i) \times Sev(F_i) = \Pr(F_i) \times \Pr(U_i | F_i) \times Sev(F_i) \qquad (1)$$

$$R_A = \frac{\sum_{i=1}^{N} R_i}{N} = \frac{\sum_{i=1}^{N} \Pr(F_i) \times \Pr(U_i | F_i) \times Sev(F_i)}{N} \qquad (2)$$

where $N$ denotes the number of MC samples (each sample represents a faulted line).

The term $\Pr(U_i \cap F_i)$ represents the joint probability of: (i) occurrence of $F_i$ ($i^{th}$ fault event), and (ii) transient instability event $U_i$. According to conditional probability theory, this term can be written as $\Pr(U_i \cap F_i) = \Pr(U_i | F_i) \times \Pr(F_i)$, as reflected by (1).

$\Pr(U_i | F_i)$ is the probability of transient instability given $F_i$ has occurred. Its value is 1 and 0 if the system is unstable and stable (for $i^{th}$ fault event), respectively [80], i.e.,

$$\Pr(U_i | F_i) = \begin{cases} 1, & \text{for } \delta_{\max i} > 360 \\ 0, & \text{otherwise} \end{cases} \qquad (3)$$

$\Pr(F_i)$ is the probability of $F_i$ ($i^{th}$ fault event) and can be defined mathematically as follows.

$$\Pr(F_i) = \Pr(F_{oi}) \times \Pr(F_{Li}) \times \Pr(F_{Ti}) \qquad (4)$$

where $\Pr(F_{oi})$, $\Pr(F_{Li})$, and $\Pr(F_{Ti})$ denote the probability of fault occurrence, fault location, and fault type, respectively, for the $i^{th}$ MC sample.

Let $F_{oi}$ be a random variable following a uniform (PMF) [81] on the interval $\{1,2,3,\ldots N_L\}$. Then,

$$\Pr(F_{oi}) = \begin{cases} \dfrac{1}{N_L}, & \text{for } 1 \leq i \leq N_L \\ 0, & \text{otherwise} \end{cases} \qquad (5)$$

where $N_L$ denotes total number of lines in the test system. Let $F_{Li}$ be a random variable following a continuous uniform PDF [81] on the interval [0,100]. Then,

$$\Pr(F_{Li}) = \begin{cases} \dfrac{1}{N_p}, & \text{for } 0 \leq i \leq N_p \\ 0, & \text{otherwise} \end{cases} \qquad (6)$$

where $N_p=100$. $\Pr(F_{Ti})$ is chosen based on PMF as shown in Table 1 [82], where $x=1, 2, 3$, and 4 denote single line to ground (LG), double line to ground (LLG), line to line (LL), and three-phase (LLL) fault, respectively.

Table 1. Probability of fault types

| $x$ | 1 | 2 | 3 | 4 |
|---|---|---|---|---|
| $\Pr(F_{Ti})$ | 0.7 | 0.15 | 0.1 | 0.05 |

$Sev(F_i)$ quantifies the impact (severity) of $F_i$. Mathematically, it is given as follows

$$Sev(F_i) = \begin{cases} |TSI_i|, & \text{if } TSI_i < 0 \\ 0, & \text{if } TSI_i > 0 \end{cases}, \quad 0 < |TSI_i| < 1 \qquad (7)$$

where $TSI_i$ denotes the transient stability index for the $i^{th}$ MC sample, i.e.,

$$TSI_i = \frac{360 - \delta_{max\ i}}{360 + \delta_{max\ i}} \qquad (8)$$

where $\delta_{max\ i}$ is the post-fault maximum rotor angle difference (in degrees) between any two SGs in the system at the same time for a fault on $i^{th}$ line [83]. A negative $TSI_i$ indicates the system in transiently unstable for the $i^{th}$ MC sample.

It is appropriate to model the uncertainty for each bus load forecast with a normal PDF having a mean equal to the forecasted value and an associated standard deviation [84-85]. Let $f(X_i)$ denote the PDF for load at $i^{th}$ bus, i.e.,

$$f(X_i) = \frac{1}{\sqrt{2\pi\sigma_i^2}} e^{\frac{-(X_i-\mu_i)^2}{2\sigma_i^2}} \tag{9}$$

where $\mu_i$ and $\sigma_i$ denotes the mean and standard deviation (10% of the mean) of the forecasted peak load for $i^{th}$ bus, respectively. Thus, the PDF for system load, $f(X_j)$, is given by

$$f(X_j) = \frac{1}{\sqrt{2\pi\sigma_j^2}} e^{\frac{-(X_j-\mu_j)^2}{2\sigma_j^2}} \tag{10}$$

where $\mu_j$ and $\sigma_j$ denotes the mean (259 MW) and standard deviation (11.5 MW) of the forecasted system peak load (the sum of multiple independent normally distributed random variables is normal, with its mean being the sum of the individual means, and its variance being the sum of the individual variances, i.e., the square of the standard deviation is the sum of the squares of the individual standard deviations), respectively.

The FCT is assumed to follow a normal PDF [86-87], with a mean and standard deviation of 0.9 *s* and 0.1 *s*, respectively.

Let $P_{LGI}$, $P_{LLI}$, $P_{LLGI}$, and $P_{LLLI}$ denote the probability of instability for LG, LL, LLG, and LLL faults, respectively. Mathematically,

$$P_{LGI} = \frac{N_{u1}}{N} \tag{11}$$

$$P_{LLI} = \frac{N_{u2}}{N} \tag{12}$$

$$P_{LLGI} = \frac{N_{u3}}{N} \tag{13}$$

$$P_{LLLI} = \frac{N_{u4}}{N} \tag{14}$$

where $N_{u1}, N_{u2}, N_{u3}$, and $N_{u4}$ denote number of unstable samples for LG, LL, LLG and LLL faults, respectively.

It should be noted that in the above-mentioned formulation, it is assumed that all CBs have same level of difficulty for opening current and equal probability failures; therefore, this is not taken into account for risk assessment model.

## 4. Computation Procedure

This section elaborates the approach to rank individual lines' CBs based on the value of $R_A$, for both line and bus faults. This will enable the planners to be extra cautious towards the critical lines/buses and corresponding CBs, for improved decision making. The approach is elaborated for line and bus faults, in Figure 1 and Figure 2, respectively. For each line/bus, 2401 MC samples (using Cochran's formula, and assuming a 95% confidence level and 2% margin of error) were used to compute the value of $R_A$, using time-domain simulation, considering PDFs of various uncertainties (system load, fault type, FCT, etc.). Consequently, based on the value of $R_A$ obtained for each line/bus, the CBs were ranked. Similarly, the approach is elaborated using only three phase bus faults in Figure 3. Only single contingencies (*N*-1) are considered in this paper.

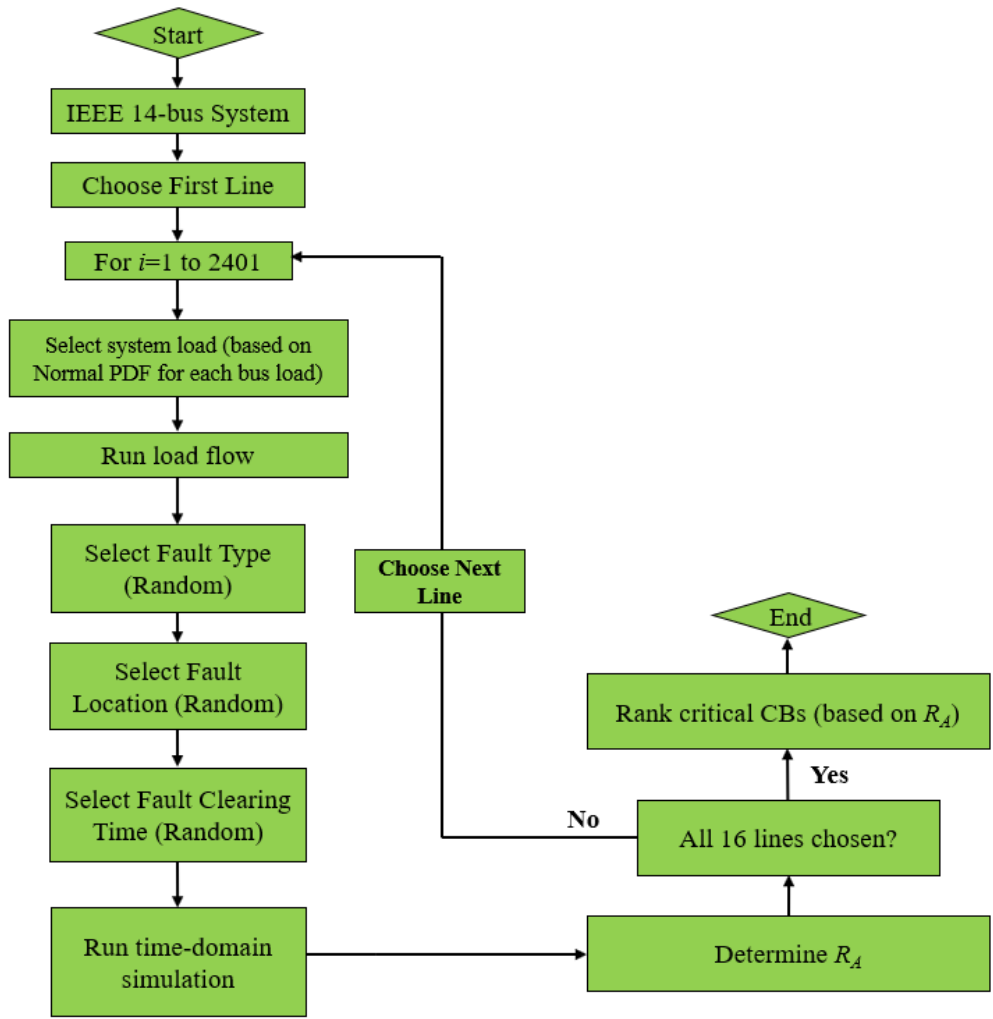

Figure 1. Methodology for CB ranking for line faults

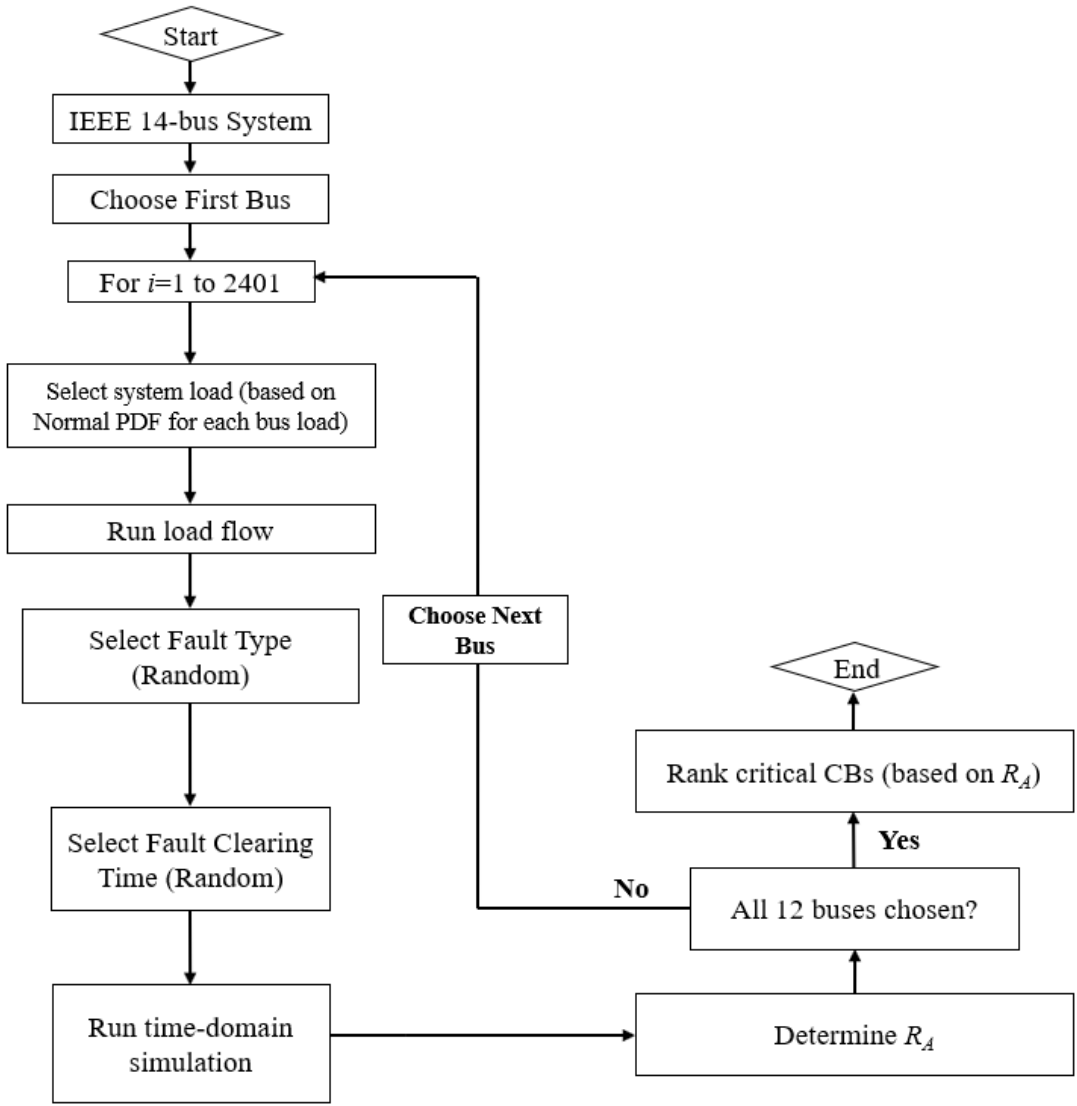

Figure 2. Methodology for CB ranking for bus faults

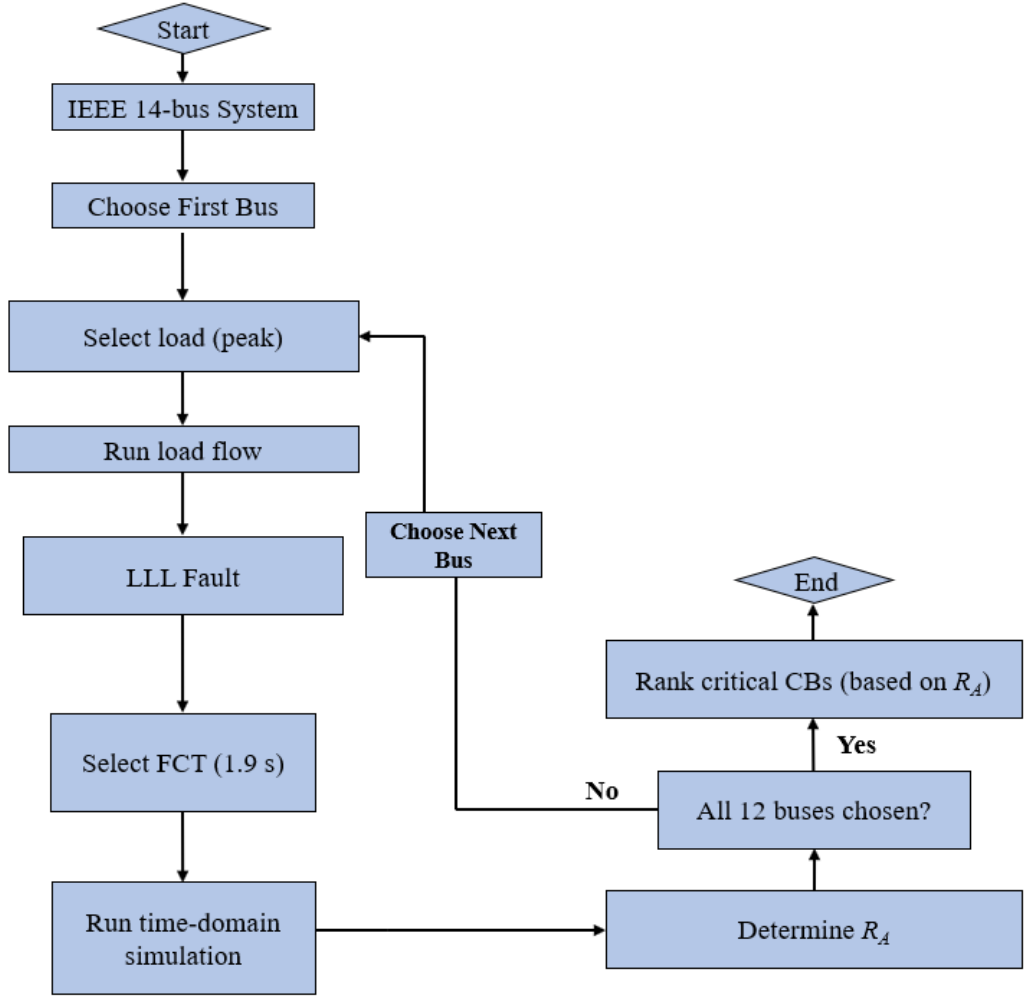

Figure 3. Methodology for CB ranking for LLL bus faults

## 5. Case Study

The IEEE 14-bus test transmission system, as shown in Figure 4, was used to conduct the required simulations. Figure 5 shows the same system with CBs labelled for each line (which will be useful for ranking CBs). It represents a simplified model of the transmission system in the Midwest United States. It consists of five synchronous machines, three of which are synchronous compensators used only for reactive power support. There are 11 loads in the system totaling 259 MW and 81.3 MVAr. This system has 16 transmission lines. The numerical data and parameters were taken from [88]. This system is a good choice for the present study, as it has been widely used by various researchers for studying transient stability phenomenon in power transmission systems [89-92]. As mentioned before, a normal (Gaussian) PDF is used to define the uncertainty in system loads. The active power of each load was assigned a mean equal to the original load active power value, as given in test system data in [88], and a standard deviation equal to 10% of the mean value. All time-domain simulations are RMS simulations and were performed using DIgSILENT PowerFactory software [93].

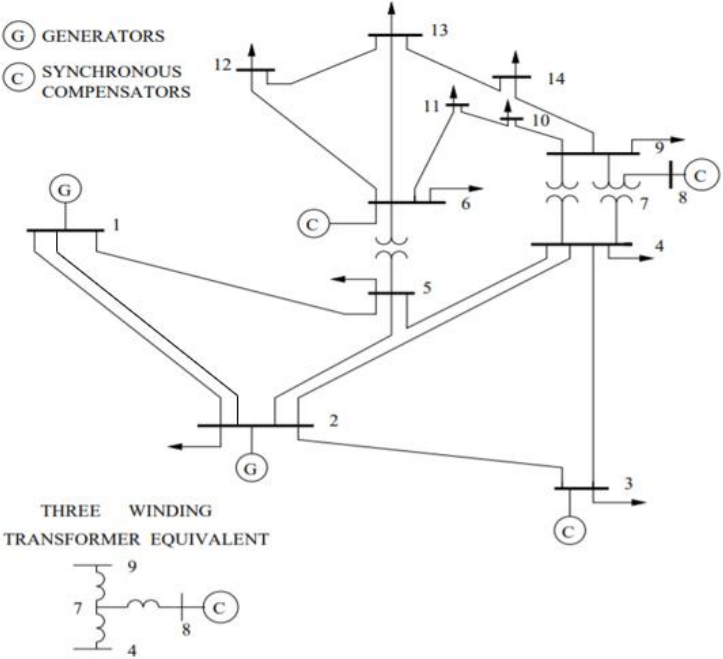

Figure 4. IEEE 14-bus system

Figure 5. IEEE 14-bus system (with CB labels)

## 6. Results and Discussion

Based on the approach discussed in Section 4, the CBs were first ranked for line faults, as shown in Table 2 (red indicates highest priority and yellow indicates lowest priority) and Figure 6. As evident, CBs B1 and B2 (on line_0006_0013) are the most critical, and CBs B31 and B32 (on line_0009_0010) are the least critical. Similarly, the CBs were ranked for bus faults, as shown in Table 3 (red indicates highest priority and yellow indicates lowest priority) and Figure 7. It must be noted that for bus faults, the following probabilities were assumed: $P_{LG}$=0.007, $P_{LLG}$=0.0015, $P_{LL}$=0.001, $P_{LLL}$=0.0005.

As evident, CBs B1, B9, and B15 (associated with bus_0006) are the most critical, and CBs B30 and B31 (associated with bus_0010) are the least critical. This quantification is essential to power planners, as it can identify which lines/buses require more attention while planning a system. These results can aid in efficient decision making, considering the level of risk associated with CBs. The index $R_A$ can be useful in devising appropriate preventive replacement and maintenance strategies for CBs and their health diagnostics. The value of this index can lead to informed decisions by power planners and operators such that the power system is always stable. This is essentially a mathematical optimization problem, and its solution can be achieved by incorporating the risk index in probabilistic cost benefit analysis using MC simulation. System planners can then use risk assessment to make decisions about breaker maintenance and replacement strategies by evaluating several key factors such as failure probability, criticality assessment, life cycle cost, and scenario analysis.

The CBs were also ranked using deterministic three phase bus faults (including other deterministic factors, such as load, FCT, etc.). The results are shown in as shown in Table 4 (red indicates highest priority and yellow indicates lowest priority) and Figure 8. As evident, CBs B1, B9, and B15 (associated with bus_0006) are the most critical, and CBs B30 and B31 (associated with bus_0010) are the least critical. Moreover, as expected, the value of $R_A$ is significantly higher for the deterministic case (only three phase bus faults) as compared to the probabilistic case. This is because for the deterministic case (probability of LLL fault being 1), the impact is greater as compared to probabilistic case (which includes all faults).

Recent research [94-102] strongly indicates that risk-based transient stability and security in power systems is a growing area, and in-depth research is required to further expand its horizon, especially with the rising uncertainties in the domain of power system protection (including CBs).

Table 2. Ranking of CBs (based on $R_A$) for line faults

| Priority Rank Index | Line | CBs | $R_A$ (%) |
|---|---|---|---|
| 1 | Line_0006_0013 | B1, B2 | 0.0058 |
| 2 | Line_0009_0014 | B3, B4 | 0.0056 |
| 3 | Line_0012_0013 | B5, B6 | 0.0056 |
| 4 | Line_0002_0005 | B7, B8 | 0.0054 |
| 5 | Line_0006_0011 | B9, B10 | 0.0053 |
| 6 | Line_0001_0002/2 | B11, B12 | 0.0052 |
| 7 | Line_0001_0002/1 | B13, B14 | 0.0051 |
| 8 | Line_0006_0012 | B15, B16 | 0.005 |
| 9 | Line_0002_0004 | B17, B18 | 0.0049 |
| 10 | Line_0013_0014 | B19, B20 | 0.0048 |
| 11 | Line_0003_0004 | B21, B22 | 0.0047 |
| 12 | Line_0004_0005 | B23, B24 | 0.0045 |
| 13 | Line_0002_0003 | B25, B26 | 0.0042 |
| 14 | Line_0001_0005 | B27, B28 | 0.0041 |
| 15 | Line_0010_0011 | B29, B30 | 0.0039 |
| 16 | Line_0009_0010 | B31, B32 | 0.0037 |

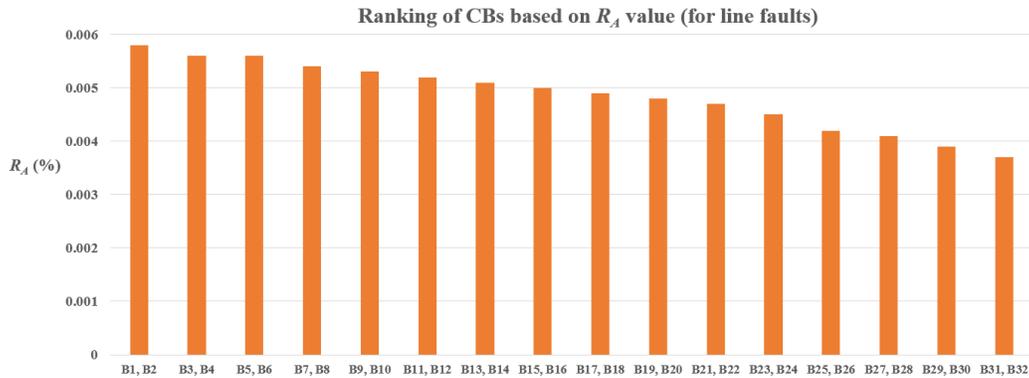

Figure 6. Ranking of CBs (based on $R_A$) for line faults

Table 3. Ranking of CBs (based on $R_A$) for bus faults

| Priority Rank Index | Bus | CBs | $R_A$ (%) |
|---|---|---|---|
| 1 | Bus_0006 | B1, B9, B15 | 0.0061 |
| 2 | Bus_0009 | B3, B32 | 0.0059 |
| 3 | Bus_00012 | B5, B16 | 0.0058 |
| 4 | Bus_0005 | B8, B18, B23 | 0.0056 |
| 5 | Bus_00011 | B10, B29 | 0.0055 |
| 6 | Bus_0001 | B11, B13, B27 | 0.0054 |
| 7 | Bus_0002 | B7, B12, B14, B17 | 0.0054 |
| 8 | Bus_0004 | B18, B22, B24 | 0.0052 |
| 9 | Bus_0013 | B2, B6, B19 | 0.005 |
| 10 | Bus_0014 | B4, B20 | 0.0048 |
| 11 | Bus_0003 | B21, B26 | 0.0046 |
| 12 | Bus_0010 | B30, B31 | 0.0044 |

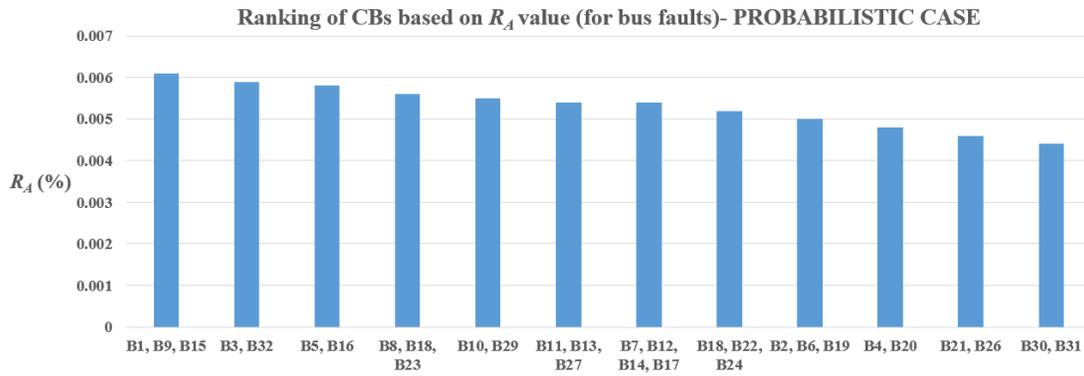

Figure 7. Ranking of CBs (based on $R_A$) for bus faults

Table 4. Ranking of CBs (based on $R_A$) for LLL bus faults

| Priority Rank Index | Bus | CBs | $R_A$ (%) |
|---|---|---|---|
| 1 | Bus_0006 | B1, B9, B15 | 1.02 |
| 2 | Bus_0009 | B3, B32 | 0.96 |
| 3 | Bus_00012 | B5, B16 | 0.93 |
| 4 | Bus_0005 | B8, B18, B23 | 0.88 |
| 5 | Bus_00011 | B10, B29 | 0.84 |
| 6 | Bus_0001 | B11, B13, B27 | 0.81 |
| 7 | Bus_0002 | B7, B12, B14, B17 | 0.78 |
| 8 | Bus_0004 | B18, B22, B24 | 0.76 |
| 9 | Bus_0013 | B2, B6, B19 | 0.74 |
| 10 | Bus_0014 | B4, B20 | 0.71 |
| 11 | Bus_0003 | B21, B26 | 0.62 |
| 12 | Bus_0010 | B30, B31 | 0.61 |

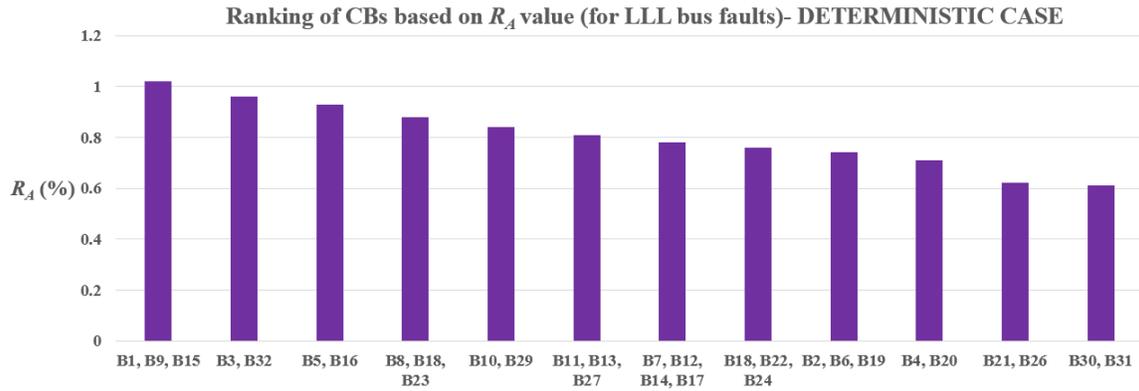

Figure 8. Ranking of CBs (based on $R_A$) for LLL bus faults

## 7. Conclusion and Future Work

Power system transient stability is an integral part of power system planning and operation. Traditionally, it has been assessed using deterministic approach. With the increasing system uncertainties, environmental pressures of incorporating green energy, and widespread electricity market liberalization (deregulation), there is a strong need to incorporate probabilistic analysis in transient stability evaluation. Therefore, this paper proposed a risk-based probabilistic transient stability approach for ranking of CBs in a power system. A novel priority index was proposed to rank the CBs based on the system transient stability risk values. Considering the results obtained for line faults, CBs B1 and B2 (on line_0006_0013) were the most critical, and CBs B31 and B32 (on line_0009_0010) were the least critical. Considering the results obtained for bus faults, CBs B1, B9, and B15 (associated with bus_0006) were the most critical, and CBs B30 and B31 (associated with bus_0010) were the least critical. The CBs were also ranked using deterministic three phase bus faults (including other deterministic factors, such as load, FCT, etc.). In this case, CBs B1, B9, and B15 (associated with bus_0006) were the most critical, and CBs B30 and B31 (associated with bus_0010) were the least critical. Moreover, the value of $R_A$ was significantly higher for the deterministic case (only three phase bus faults) as compared to the probabilistic case. DIgSILENT PowerFactory software was used to conduct the required simulations on the IEEE 14 bus system. The proposed method successfully ranks the CBs based on their priority index. The proposed index can be very helpful in maintenance, replacement, and upgradation of existing CBs.

As the proposed approach is universal and generic; hence, it can easily be extended to a large-scale system. With the quick development of natural gas fired units worldwide, the interdependency of natural gas system and power system has substantially augmented. Thus, one of the future works could include proposing a unified approach for incorporating natural gas systems in the transient stability analysis. Uncertainties introduced by unforeseen high-risk events, such as natural disasters (hurricanes, earthquakes, floods, etc.), extreme weather (ice storms, heat waves, high winds, etc.), cyber-attacks, etc. can prove lethal to the power system, and hence, can have a detrimental impact on the transient stability. The central challenge in this regard is to model these phenomena and their allied impact accurately. Consequently, CBs can be ranked incorporating these phenomena in a risk-based transient instability formulation. Also, the proposed risk assessment model can be enhanced by considering the varying difficulty of opening currents across different breaker locations and their varying probability failures. Other factors which greatly influence CB reliability such as installation quality, maintenance practices, and power system configuration, can also be incorporated. Risk assessment model can be extended to include CB interval probabilities which can greatly assist the work of power system experts and utilities and give a rational picture of the actions and decisions to be taken. This can substantially further aid in assessing the state of the system leading to accurate and timely decision-making.

**Disclosure statement**

No potential conflict of interest was reported by the author.

**Data availability statement**

Data available on request from the author.

**Funding statement**

No funding was received.

**Notes on Contributor**

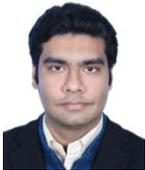 **Umair Shahzad** is currently working as a Lecturer at the School of Computer Science and Engineering, University of Sunderland, UK. In 2021, he received the Ph.D. degree in Electrical Engineering from The University of Nebraska-Lincoln, USA. Moreover, he received a B.Sc. Electrical Engineering degree from the University of Engineering and Technology, Lahore, Pakistan, and a M.Sc. Electrical Engineering degree from The University of Nottingham, England, in 2010 and 2012, respectively. His research interests include power system security assessment, power system stability, artificial intelligence, machine learning, and probabilistic methods applied to power systems.